\documentclass[]{aa}
\usepackage{psfrag}
\usepackage{graphicx}
\begin{document}
   \title{Multi-wavelength afterglow observations of the high redshift GRB~ 050730 
  }

\author{S. B. Pandey\inst{1}, A.J. Castro-Tirado\inst{1}, S. McBreen\inst{2,3}, 
M. D. P\'erez-Ram\'{\i}rez\inst{4}, M. Bremer\inst{5}, M. A. Guerrero\inst{1},
A. Sota\inst{6}, B. E. Cobb\inst{7}, M. Jel\'inek\inst{1}, 
A. de Ugarte Postigo\inst{1}, J. Gorosabel\inst{1}, S. Guziy\inst{1, 15}, C. Guidorzi\inst{9}, 
C. D. Bailyn\inst{7}, T. Mu\~noz-Darias\inst{8}, A. Gomboc\inst{9, 10}, 
A. Monfardini\inst{9}, C. G. Mundell\inst{9}, N. Tanvir\inst{11}, A. J. Levan\inst{11},
B. C. Bhatt\inst{12,13}, D. K. Sahu\inst{12,13}, S. Sharma\inst{14}, 
O. Bogdanov\inst{15} and J. A. Combi\inst{4}}

   \institute{ 
   Instituto de Astrof\'isica de Andaluc\'ia, P.O. Box 03004, E-18080, Granada, Spain
   \and
   European Space Agency, ESTEC, Keplerlaan 1, 2201 AZ Noordwijk, The Netherlands
    \and  Max-Planck-Institut
  f\"{u}r extraterrestrische Physik, 85748 Garching, Germany.
   \and
   Departamento de F\'isica (EPS), Universidad de Ja\'en, Campus Las Lagunillas s/n (Ed-A3), 23071 
  Ja\'en, Spain
   \and
   Institut de Radioastronomie Millim\'etrique, 300 rue de la Piscine, 38406 Saint Martin d'H\`eres, France
   \and
   IAA-CSIC and  Space Telescope Science Institute, St. Martin Dr., Baltimore, MA, USA
   \and
   Department of Astronomy, Yale University, P.O. Box 208101, New Haven, CT 06520, USA
   \and
   Instituto de Astrof\'{\i}sica de Canarias, C/. V\'{\i}a L\'actea, s/n, E-38200 La Laguna, 
   Tenerife, Spain
   \and
   Astrophysics Research Institute, Liverpool John Moores University, Twelve Quays House, 
   Egerton Wharf, Birkenhead, CH41 1LD, UK
   \and
   Faculty of Mathematics and Physics, University of Ljubljana, Jadranska 19, 
   1000 Ljubljana, Slovenia
   \and
   Center for Astrophysics Research, University of Hertfordshire, College Lane, Hatfield
   AL10 9AB, UK
   \and
   Center for Research \& Education in Science \& Technology, Hosakote, Bangalore -- 562 114, India
   \and
   Indian Institute of Astrophysics, Bangalore -- 560 034, India
   \and
   Aryabhatta Research Institute of Observational Sciences (ARIES), Manora Peak, Naini Tal, 
   Uttaranchal, 263129, India
   \and
   Nikolaev State University, Nikolska 24, Nikolaev, 54030, Ukraine
}

   \authorrunning{Pandey et al.}
   \titlerunning{Multi-wavelength afterglow observations of GRB~050730}
   \offprints{S. B. Pandey, \\
           \email{sbp2@mssl.ucl.ac.uk}
            }
   \date{Received ------ /accepted ---------}
\abstract
{GRB~050730 is a long duration high-redshift burst (z=3.967) discovered by \textit{Swift}. The 
afterglow shows variability and is well monitored over a wide wavelength range.
We present comprehensive temporal and spectral analysis of the afterglow of GRB~050730 
including observations from the millimeter to X-rays. We use multi-wavelength afterglow data to understand the temporal and spectral decay 
properties with superimposed variability of this high redshift burst.
Five telescopes were used to study the decaying afterglow of GRB~050730 in the 
$\rm B, V, r', R, i', I, J$ and $K$ photometric pass bands. A spectral energy distribution was 
constructed at 2.9~hours post-burst in the $\rm K, J, I, R, V$ and $\rm B$ bands.
X-ray data from the satellites \textit{Swift} and \emph{XMM-Newton} were used to study the 
afterglow evolution at higher energies.
The early afterglow shows variability at early times and shows a steepening at 0.1~days (8.6~ks) 
in the $\rm B, V, r', R, i', I, J$ and $K$ passbands. The early afterglow light curve decayed with 
$\alpha_1 = -0.60\pm0.07$ and $\alpha_2 = -1.71\pm0.06$ based on $\rm R$ and $\rm I$ band data.
A millimeter detection of the afterglow around 3 days after the burst shows
an excess in comparison to predictions. The early X-ray light curve observed by \textit{Swift}
is complex and contains flares. At late times the X-ray light curve can be fit by a powerlaw 
$\alpha_x = -2.5\pm0.15$ which is steeper than the optical light curve.
A spectral energy distribution (SED) was constructed at $\sim$2.9 hours after the
burst. An electron energy index, p, of $\sim$2.3 was calculated using the SED and
the photon index from the X-ray afterglow spectra and indicates that the
synchrotron cooling frequency $\nu_c$ is above observed frequencies.
}

\maketitle
\section{Introduction}
GRB~050730 is one of  a growing number of known cases at high redshift for 
which peculiar superimposed variability is observed from optical to X-rays 
(e.g. GRB050904: B\"oer et al. 2005; Cusumano et al. 2006; Watson et al. 2005). Observed variability contains a wealth of information about the nature of the 
burst mechanism (Burrows et al. 2005). The very early X-ray and optical 
observations of GRB~050730 were possible due to the very fast slew time of 
{\textit Swift} satellite (Gehrels et al 2004).
  
Long-duration GRB~050730 was detected by {\it Swift}-BAT (trigger=148225) at 
$T_0$ = 19$^{\rm h}$58$^{\rm m}$23$^{\rm s}$ UT on 30th July 2005 
(Holland et al. 2005). The X-ray
and the optical afterglow of the burst were discovered by the on-board
instruments XRT and UVOT respectively, after 132s and 119s after the 
BAT trigger (Holland et al. 2005). The optical afterglow (OA) candidate was later 
confirmed by ground based observations using the Sierra Nevada 1.5m telescope
by Sota et al. (2005). The near infra-red ($\rm NIR$) afterglow was discovered 
by Cobb \& Bailyn (2005) using ANDICAM mounted at the CTIO 1.3m. Spectroscopic 
observations of the afterglow candidate, obtained $\sim$ 4 hours after the burst, 
using the MIKE echelle 
spectrograph on Magellan II derived a redshift of z = 3.967 (Chen et al. 2005a).
The redshift value was further confirmed by Rol et al. (2005) and Prochaska
et al. (2005). The derived redshift value was based on the strong absorption
feature at $\lambda$ 6040\AA\, identified as a hydrogen Ly-$\alpha$ together with 
other narrow absorption lines due to heavy ions originating in the surroundings of 
the GRB progenitor (Chen et al.\ 2005a,b).    

The source was initially detected by {\textit Swift}/BAT at
RA(J2000)=14$^{\rm h}$08$^{\rm m}$16.40$^{\rm s}$, 
Dec(J2000)=--03$^\circ$45$^\prime$41\farcs1 with an uncertainty of 
3$^{\prime}$.  Markwardt et al. (2005) reported the burst 
duration (T$_{90}$) 155$\pm$20\,s and fluence 
(4.4$\pm$0.4)$\times 10^{-6}$erg cm$^{-2}$ s$^{-1}$.
Initial analysis of  XRT data (130 to 1000 seconds 
after the burst) show flaring in the light-curve (Grupe, 
Kennea \& Burrows 2005; Perri et al. 2005; Starling et al. 2005). {\textit Swift}/XRT 
error box was also observed by \emph{XMM-Newton}
(Schartel 2005), confirming the presence of the 
afterglow of GRB~050730 (Ehle \& Juarez 2005). VLA radio observations around 
3 days after the burst show a weak radio source consistent with the optical 
afterglow candidate at 8.5 GHz (Cameron 2005). Three WSRT observations 
at 4.9 GHz (van der Horst \& Rol 2005 a,b) at $\sim$4.6, 6.6 and 12.6 days 
postburst show no significant flux at the radio afterglow position.      

This paper is organized as follows. Afterglow observations spanning 
a wide wavelength range, including millimeter, near-infrared (NIR), optical and 
X-rays, are described in Section 2. The results of the multi-wavelength 
analysis are presented in Section 3. A detailed discussion of the afterglow 
light-curves, spectral energy distribution and their comparison to model 
predictions can be found in Section 4. Finally, concluding remarks are 
addressed in Section 5.
 
\section {GRB~050730 afterglow Observations}  
In the following sections we describe the observations, subsequent
data reduction and calibration of the afterglow of GRB~050730 at 
millimeter, NIR, optical and X-ray wavelengths.

\subsection{Millimeter Wave Observations }

Observations were triggered at the Plateau de Bure Interferometer 
(PdBI, Guilloteau et al. 1992) as part of an on-going Target of Opportunity (ToO)
programme. The observations were centered on the equatorial coordinates
RA(J2000)=14$^{\rm h}$08$^{\rm m}$17.14$^{\rm s}$, Dec(J2000)=--03$^\circ$46$^\prime$17\farcs8. 
The counterpart was observed on August 2 and 5 2005 in a compact 5 antenna configuration (5D) 
and on January 5 2006 in a 6 antenna configuration (6Cp). The later observation in January 
was designed to assess the influence of the underlying host galaxy on the initial 
afterglow observations. Standard software packages,  CLIC and MAPPING,
distributed by the Grenoble GILDAS group\footnote{GILDAS is a software package 
distributed by the IRAM Grenoble GILDAS group.} were used to reduce the 
data. Flux calibration was based on the carbon-star MWC349. The flux values and 
error estimates were established with point source fits in the UV plane, which were
fixed to the phase center coordinates.

The afterglow continuum was tentatively detected in the 3mm band on
August 2 and 5 at 2.9 and $2.6 \sigma$ respectively
(Table 1). A combination of both data sets (with the appropriate
weighting for system temperature and amplitude calibration) gave a
flux of $1.73 \pm 0.45$~mJy, i.e. a $3.8 \sigma$ detection on the
phase center. Only upper limits could be obtained 
in the 1mm band.

A final observation on January 5th, 2006 under good atmospheric
conditions provided upper limits in both 3mm and 1mm bands, 
indicating that the contribution of the host galaxy is not significant.

\begin{table*}[htdp]
{\bf Table 1.}~ PdBI dual frequency observations of GRB~050730. First three
lines: 3mm band, last three lines: 1mm band. The flux errors are
$1\sigma$. Last two columns represent the beam size and position angle (PA)
respectively.  
\begin{center}
\begin{tabular}{cccccl}
\hline
\hline
Start time & End time & Freq. [GHz] & Flux [mJy] & Beam($^{\prime\prime} \times ^{\prime\prime}$) & PA ($^\circ$) \\
\hline \hline
2005 Aug 2.659 & 2.816 & 102.746 & $ 2.74\pm 0.94$ & 9.2$\times$ 5.7 & 49 \\
2005 Aug 5.577 & 5.838 & 105.304 & $ 1.34\pm 0.51$ & 6.9$\times$ 6.4 & 73 \\
2006 Jan 5.193 & 5.356 &  86.847 & $-0.24\pm 0.27$ & 4.4$\times$ 2.9 & 4  \\
\hline
2005 Aug 2.659 & 2.816 & 213.233 & $ 4.39 \pm 4.21$ & 4.8$\times$ 2.5 & 10 \\
2005 Aug 5.577 & 5.838 & 214.977 & $-5.13 \pm 3.81$ & 3.8$\times$ 2.9 & 0  \\
2006 Jan 5.193 & 5.356 & 230.538 & $-4.21 \pm 1.93$ & 1.7$\times$ 1.1 & 1  \\
\hline
\end{tabular}
\end{center}
\end{table*}

\subsection {Near$\rm -IR$ observations}  
 
The $\rm NIR$ observations of the afterglow started at 22:46 UT
on 30th July 2005 under non-photometric sky conditions using the ANDICAM 
instrument mounted on the 1.3m telescope at Cerro Tololo Inter-American Observatory
(CTIO)\footnote{http://www.astronomy.ohio-state.edu/ANDICAM}, operated as 
a part of the Small and Moderate Aperture Research Telescope System (SMARTS) 
consortium\footnote{http://www.astro.yale.edu/smarts}. The ANDICAM detector 
consists of a dual-channel camera that allows simultaneous optical ($\rm V, I$) 
and $\rm NIR$ ($\rm J, K$) imaging. $\rm NIR$ and optical images are acquired 
simultaneously by the ANDICAM instrument via an internal mirror which 
repositions the $\rm NIR$ image on the CCD, essentially "dithering"
without physically moving the telescope and without interrupting the 
optical observations.

A combination of telescope pointings and internal dithers were used to
obtain 16 $\rm J$ and $\rm K$ band images. The reduction process was as follows.  
A master dome flat was produced for each of the 4 internal dither positions and 
the images were divided by the relevant master field flat. A sky frame in the 
$\rm J$ and $\rm K$ band for each dither position was obtained by the median 
combination of 4 images. Sky frames were subtracted 
from each image with rescaling to compensate for changes in brightness. The field was 
re-observed on subsequent photometric nights for calibration purposes. 
Standard fields Persson-P9104 and LCO-BRI0021 (Persson et al. 1998) were used for $\rm NIR$ 
calibrations. The 16 resulting $\rm J$ and $\rm K$ band images were combined into sets of
4 or 8 to search for variability of the afterglow. A log of the $\rm J$ and $\rm K$ band 
observations on performed on the 30 and 31 July 2005 can be found in Table 2.

Four 225~s $\rm I$ and $\rm V$-band images were also obtained by 
ANDICAM on 30 July and 1 August 2005. Standard reduction was performed on the 
optical images, including over-scan bias subtraction, zero subtraction and flat 
fielding as discussed in the following subsection. 
The log of these observations is also presented in Table 2.

\subsection {Optical observations}  

Broad-band observations of the optical afterglow in the Bessel $\rm B,V, R, I$
bands were carried out at various epochs between 30 July to 01 
August 2005 using the 2-m Himalayan Chandra Telescope (HCT) of the Indian 
Astronomical Observatory (IAO, Hanle India), the 1.5m telescope at Observatorio de 
Sierra Nevada (OSN) in Granada (Spain), the 1.3m CTIO on Kitt Peak Arizona USA and 
the Instituto de Astrof\'{\i}sica de Canarias (IAC) 0.8-m telescope at 
Observatorio de Iza\~na in Tenerife, Spain. Bessel $\rm B, V$ and $\rm SDSS$ 
$r', i'$ observations of the OA were obtained on 30 July 2005 using the 
2-m robotic Liverpool Telescope (LT) of John Moores University at Canary 
Islands (La Palma, Spain).
Several twilight flat and bias frames were also obtained during the 
observing runs for the CCD calibrations. In order to improve the 
signal-to-noise ratio of the OA the data have been binned in $2 \times 2$ 
pixel and images were co-added when necessary. 
Profile fitting magnitudes were determined from the images using
a standard procedure in DAOPHOT/IRAF\footnote{http://iraf.noao.edu/}.

\begin{table}[h]
{\bf Table 2.}~Observational log of the photometric CCD magnitudes in 
Bessell $\rm B, V, R, I$, SDSS $r', i'$ broad-band optical and $J, K$ $\rm NIR$ 
observations of the GRB~050730 afterglow. 

\begin{center}
\scriptsize
\begin{tabular}{ccll} \hline \hline 
Date (UT) of & Magnitude & Exposure time & Telescope  \\
2005 & &(Seconds)&  \\   \hline \hline 
\multicolumn{3}{c}{\bf $B-$ passband}  \\
July 30.8608&19.68$\pm$0.04&300 &OSN 1.5m \\
July 30.8745&19.61$\pm$0.03&300 &OSN 1.5m \\ 
July 30.8899&19.77$\pm$0.04&300 &OSN 1.5m \\ 
July 30.9567&19.96$\pm$0.18&240 &LT 2.0m \\ 
 \multicolumn{3}{c}{\bf $V-$ passband} \\ 
July 30.8572&18.00$\pm$0.01&300 &OSN 1.5m \\ 
July 30.8718&18.21$\pm$0.01&300 &OSN 1.5m \\ 
July 30.8863&18.28$\pm$0.02&300 &OSN 1.5m \\ 
July 30.9018&18.60$\pm$0.02&300 &OSN 1.5m \\ 
July 30.9162&18.69$\pm$0.04&300 &OSN 1.5m \\ 
July 30.9528&18.56$\pm$0.08&240 &LT 2.0m \\ 
July 30.9536&18.67$\pm$0.12&225 &CTIO 1.3m \\ 
July 30.9604&18.92$\pm$0.08&225 &CTIO 1.3m \\ 
July 30.9623&18.92$\pm$0.06&225 &CTIO 1.3m \\ 
July 30.9638&18.92$\pm$0.07&225 &CTIO 1.3m \\ 
July 30.9711&19.13$\pm$0.07&225 &CTIO 1.3m \\ 
 \multicolumn{3}{c}{\bf $r'-$ passband} \\ 
July 30.8625&16.78$\pm$0.20&120 &LT 2.0m \\ 
July 30.8670&17.11$\pm$0.10&240 &LT 2.0m \\ 
July 30.8747&17.29$\pm$0.05&360 &LT 2.0m \\ 
July 30.8841&17.46$\pm$0.04&240 &LT 2.0m \\ 
July 30.8872&17.50$\pm$0.04&240 &LT 2.0m \\ 
July 30.8991&17.74$\pm$0.04&240 &LT 2.0m \\ 
July 30.9022&17.84$\pm$0.05&240 &LT 2.0m \\ 
July 30.9426&17.70$\pm$0.05&240 &LT 2.0m \\ 
July 30.9604&17.88$\pm$0.04&240 &LT 2.0m \\ 
 \multicolumn{3}{c}{\bf $R-$ passband} \\ 
July 30.8536&17.07$\pm$0.01&300 &OSN 1.5m \\ 
July 30.8682&17.46$\pm$0.01&300 &OSN 1.5m \\ 
July 30.8827&17.42$\pm$0.01&300 &OSN 1.5m \\ 
July 30.8982&17.78$\pm$0.01&300 &OSN 1.5m \\ 
July 30.9053&17.93$\pm$0.01&300 &OSN 1.5m \\ 
July 30.9126&17.88$\pm$0.02&300 &OSN 1.5m \\ 
July 30.9197&17.91$\pm$0.02&300 &OSN 1.5m \\ 
July 31.6736&21.48$\pm$0.07&300+900&HCT 2.01m \\ 
 \multicolumn{3}{c}{\bf $i'-$ passband} \\ 
July 30.8644&16.68$\pm$0.13&120 &LT 2.0m \\ 
July 30.8705&16.76$\pm$0.06&240 &LT 2.0m \\ 
July 30.8798&16.74$\pm$0.03&360 &LT 2.0m \\ 
July 30.8909&16.90$\pm$0.04&240 &LT 2.0m \\ 
July 30.8939&16.92$\pm$0.04&240 &LT 2.0m \\ 
July 30.9066&17.18$\pm$0.04&360 &LT 2.0m \\ 
July 30.9105&17.14$\pm$0.04&240 &LT 2.0m \\ 
July 30.9469&17.08$\pm$0.04&240 &LT 2.0m \\ 
 \multicolumn{3}{c}{\bf $I-$ passband} \\ 
July 30.8495&15.66$\pm$0.01&300 &OSN 1.5m \\ 
July 30.8646&16.35$\pm$0.01&300 &OSN 1.5m \\ 
July 30.8791&16.34$\pm$0.01&300 &OSN 1.5m \\ 
July 30.8941&16.59$\pm$0.01&300 &OSN 1.5m \\ 
July 30.9088&16.87$\pm$0.01&300 &OSN 1.5m \\ 
July 30.9232&16.80$\pm$0.02&300 &OSN 1.5m \\ 
July 30.9503&16.74$\pm$0.01&400 &IAC80 \\ 
July 30.9517&16.87$\pm$0.03&225 &CTIO 1.3m \\ 
July 30.9623&16.97$\pm$0.02&225 &CTIO 1.3m  \\ 
July 30.9671&17.10$\pm$0.02&225 &CTIO 1.3m  \\ 
July 30.9769&17.22$\pm$0.02&225 &CTIO 1.3m  \\ 
July 31.9167&20.75$\pm$0.08&900+1800+2200&IAC80 \\ 
Aug 01.1963&20.73$\pm$0.01&360$\times$6 &CTIO 1.3m  \\ 
Aug 03.0341&22.64$\pm$0.46&300$\times$11&OSN 1.5m \\ 
 \multicolumn{3}{c}{\bf $J-$ passband} \\ 
July 30.9537&16.17$\pm$0.07&45$\times$3&CTIO 1.3m \\ 
July 30.9605&16.14$\pm$0.08&45$\times$3&CTIO 1.3m \\ 
July 30.9639&16.55$\pm$0.08&45$\times$3&CTIO 1.3m \\
July 30.9711&16.46$\pm$0.07&45$\times$3&CTIO 1.3m \\ 
July 31.0285& $>$ 19.5$\pm$0.1&60$\times$30& CTIO 1.3m\\ 
 \multicolumn{3}{c}{\bf $K-$ passband} \\ 
July 30.9538&14.73$\pm$0.10&45$\times$8&CTIO 1.3m \\ 
July 30.9708&15.01$\pm$0.10&45$\times$8&CTIO 1.3m \\ 
\hline
\end{tabular} 
\end{center} 
\end{table} 

\normalsize

During good photometric sky conditions at 1.04m reflector Naini Tal\footnote{http://aries.ernet.in/}, 
the CCD $\rm B, V, R, I$ observations 
of the OA field and the Landolt (1992) standard PG1323-085 region were obtained 
along with several twilight flat and bias frames on 31 Dec/01 Jan 2006 for 
calibration purposes at similar airmass values. The values of atmospheric
extinction coefficients determined from Naini Tal in $B, V, R$ and $I$ filters
were 0.27, 0.17, 0.11 and 0.09 mag respectively. The observed standard stars 
in the PG1323-085 region cover a range of $-0.13 < (V-I) < 0.83$ in color and 
of $12.1 < V < 14.0$ in brightness. The zero points and the associated errors 
were determined using standard DAOPHOT/IRAF routines using nearby stars. 
The calibrated $\rm B, V, R$ and $\rm I$ 
magnitudes of 10 nearby stars to the afterglow candidate are tabulated 
in Table 3. The $\rm B, V, R, I$ magnitudes of the afterglow candidate are calibrated 
differentially with respect to these secondary standards (Table 3) and the 
magnitudes of the afterglow candidate derived in this way are given in Table 2. 

\begin{table}[h]
{\bf Table 3.}~The identification number (ID), $(\alpha , \delta)$ for epoch 2000,
standard $V, (B-V), (V-R)$ and $(V-I)$ photometric magnitudes of the stars in
the GRB~050730 region are given. 

\begin{center}
\scriptsize
\begin{tabular}{ccc cc ccl} \hline \hline
ID & RA(J2000) & Dec(J2000) & $V$& $B-V$ & $V-R$ & $V-I$ \\
    & $({\rm h:m:s})$ & $(\circ:\prime:\prime\prime)$ & (mag) & (mag) & (mag) & (mag)  \\ \hline \hline
 01&14 08 10.9&-03 46 40.2& 15.15&  0.41&  0.23&  0.73 \\
 02&14 08 14.6&-03 46 27.7& 18.06&  0.53&  0.29&  0.89 \\
 03&14 08 17.2&-03 46 36.1& 18.94&  1.10&  0.68&  1.72 \\
 04&14 08 19.4&-03 45 43.9& 19.02&  0.56&  0.35&  1.07 \\
 05&14 08 22.1&-03 48 16.9& 16.11&  1.13&  0.67&  1.61 \\
 06&14 08 18.7&-03 49 14.9& 15.70&  0.50&  0.29&  0.87 \\
 07&14 08 18.4&-03 44 46.8& 17.72&  0.76&  0.45&  1.18 \\
 08&14 08 14.5&-03 44 42.6& 17.51&  0.73&  0.46&  1.18 \\
 09&14 08 11.6&-03 44 36.8& 17.30&  0.64&  0.36&  1.01 \\
 10&14 08 26.7&-03 46 51.9& 16.56&  0.71&  0.41&  1.07 \\
\hline
\end{tabular}
\end{center}
\end{table}

\normalsize

\subsection {X-ray observations}  

\subsubsection {\textit{Swift} observations}  

{\it Swift}/XRT began to observe GRB~050730 (Trigger 148225) 132~s after the 
trigger and the data confirm the presence of a decaying X-ray source in 
the {\it Swift}/XRT field at position RA(J2000)=14$^{\rm h}$08$^{\rm m}$17.5$^{\rm s}$, 
Dec(J2000)=--03$^\circ$ 46$^\prime$19\farcs0 with an uncertainty of 
6$^{\prime\prime}$ as reported by Perri et al. (2005).
The XRT 
data consist of Window Timing (WT) data from T$_0+$132~s to T$_0+$790~s and Photon 
Counting (PC) data from T$_0+$ 4~ks and onwards. The data were reduced using the standard pipeline 
for XRT data analysis software\footnote{http://swift.gsfc.nasa.gov
/docs/software/lheasoft/download.html}(version 2.2) and the most recent 
calibration files are used. The data were analyzed with the XSPEC version 11.3 
(Arnaud 1996).

\begin{figure}[htbp]
\begin{center}
\includegraphics[width=\columnwidth]{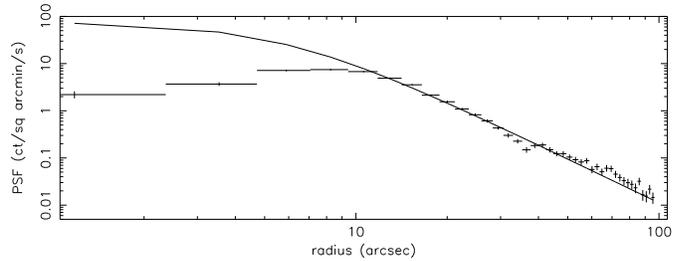}
\caption{GRB~050730 X-ray afterglow Photon Counting data for T$_0+$4~ks to 
T$_0+$6.6~ks.  The point spread function and the data diverge at $\sim$\,10 
arc sec. Only events outside the 10 arc sec region can be used for 
the analysis.
\label{pileup}}
\end{center}
\end{figure}

Source and background regions were extracted using a rectangular aperture for the 
WT data. The PC data from T$_0+$4~ks to T$_0+$24~ks was "piled-up" due to the 
intensity of the source emission. Pile-up occurs when more than one photon is 
collected in CCD frame and they are counted as a single event of higher energy. 
The main result is an apparent 
loss of flux from the center of the Point Spread Function (PSF) as shown in 
Figure\,\ref{pileup} (for a detailed discussion of pile up in XRT see Vaughan et al. 2006).
Figure\,\ref{pileup} shows the PSF of the XRT and the PC data from T$_0+$4 to 
T$_0+$6.6~ks. Clearly the count rate is diminished inside 10 arc sec and this region
of the CCD should not be used for spectra and in addition care must be
taken to estimate the true count rate. Annular source regions were used to extract 
spectra for the piled-up PC data.  The radius of the affected inner annulus was 
determined by fitting the PSF to the data and selecting regions where the data are 
well fitted by the PSF (as shown in Figure\,\ref{pileup}). Circular source regions 
were used to extract the spectra 
after 27~ks since the rate had fallen below the critical level 0.6 counts s$^{-1}$.
In addition, the data were affected by bad CCD columns and the rate was corrected 
(A. Beardmore \& K. Page, private communication).

\begin{figure}
\includegraphics[width=\columnwidth]{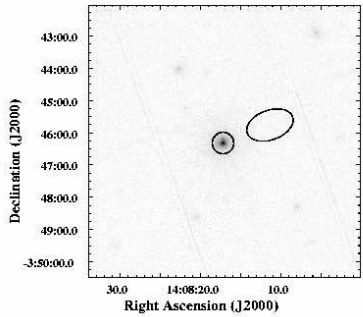}
\caption{
\emph{XMM-Newton} raw EPIC image in the 0.45-8.0 keV energy band of the 
region around the \emph{Swift}-BAT error box of GRB\,050730. The X-ray 
afterglow of GRB\,050730 is the bright central source in this image.  
The circle and ellipse indicate the spatial regions used for source 
and background extraction, respectively. 
\label{pn-image}}
\end{figure}

\subsubsection {\emph{XMM-Newton} observations}  

Target of Opportunity \emph{XMM-Newton} observations\footnote{Based on observations obtained with 
\emph{XMM-Newton}, an ESA science mission with instruments and contributions
directly funded by ESA Member States and NASA} of the region around the {\textit Swift}-BAT 
error box of GRB\,050730 were triggered in revolution 1033 (Observation ID.\ 
0164571201) (Schartel (2005)). The observations started at 
03:00:08 {\sc ut} on 2005 July 31, i.e., $\sim$25 ks after the initial burst. The EPIC/MOS1, 
EPIC/MOS2 and EPIC/pn CCD cameras were operated in the Prime Full Window Mode for a total 
exposure time of 33.2 ks for the EPIC/MOS and 25.7 ks for the EPIC/pn.  
The EPIC/MOS1 observations used the Medium optical blocking filter, 
while the EPIC/MOS2 and the EPIC/pn observations used the Thin1 
filter.  

The \emph{XMM-Newton} pipeline products were processed using the 
\emph{XMM-Newton} Science Analysis Software (SAS version 6.1.0) and 
the calibration files from the Calibration Access Layer available 
on 2005 November 24. 
Time intervals with high background (i.e., count rates in the 
background-dominated 10--12 keV energy range $\ge$0.4 cnts~s$^{-1}$ 
for the EPIC/MOS or $\ge$1.2 cnts~s$^{-1}$ for the EPIC/pn) were discarded. 
The resulting exposure times are 26.4 ks, 25.5 ks, and 17.9 ks for 
the EPIC/MOS1, EPIC/MOS2 and EPIC/pn observations respectively.

In order to search for the X-ray afterglow of GRB\,050730, we extracted 
raw \emph{XMM-Newton} EPIC/MOS1, EPIC/MOS2 and EPIC/pn images in the 
0.45-8.0 keV band with a pixel size of 2\arcsec. These images show a 
bright X-ray point-source within the \emph{Swift}-BAT error box of GRB\,050730, 
as clearly seen in the merged EPIC image shown in Figure~\ref{pn-image}.  
The location of this X-ray source, at RA(J2000)=14$^{\rm h}$8$^{\rm m}$17.2$^{\rm s}$, 
Dec(J2000)=--3$^\circ$46$^\prime$18\farcs6, is coincident with the optical 
transient of GRB\,050730 (Sota et al. 2005), thus allowing the identification
of the X-ray afterglow of GRB\,050730. The \emph{XMM-Newton} 
EPIC/MOS1, EPIC/MOS2, and EPIC/pn observations of this source detect a total of 
7,350$\pm$100 cnts, 7,700$\pm$100 cnts and 19,000$\pm$140 cnts respectively. We would 
like to emphasize that this X-ray source is placed away from the EPIC/MOS and 
EPIC/pn CCD gaps, thus allowing a reliable count rate determination.  



\section{ Results}

\subsection{ $\rm NIR$-Optical photometric light-curves}

The optical and NIR light-curves of the afterglow of GRB~050730 in the
 $\rm B, V, r', R, i', I$ and $\rm J$ passbands are shown in Figure 3. 
Observations presented in this figure are supplemented by other photometric measurements of 
the afterglow published in GCN circulars (Figure 3 and the caption).
The supplementary data were also calibrated using the secondary standards 
tabulated in Table 3.

\begin{figure}[h]
\centering
\includegraphics[width=\columnwidth]{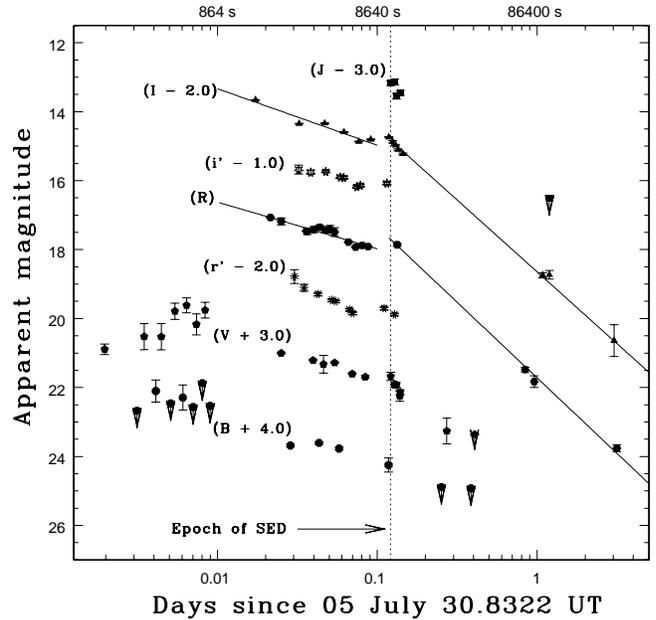}
\caption{Optical light-curves of GRB~050730 afterglow in  $\rm B, V, r', R, i', I$ 
and $\rm J$ pass bands. Marked vertical offsets are applied to avoid overlapping 
of data points in different passbands. The solid curves are the linear 
least square best fitted power-law relations for $\rm R$ and $\rm I$ pass-bands.
Vertical dotted line shows the epoch of NIR-optical SED shown in Figure 4.
Data presented in this work (Table 2)
and supplementary data (Bhatt \& Sahu 
(2005), Blustin et al. (2005), Burenin et al. (2005), Damerdji et al. (2005), 
Gomboc et al. (2005), Haislip et al. (2005), and Kannappan et al. (2005))
are combined in this figure. 
\label{optical}}
\end{figure}

The light-curve in Figure 3 is presented relative to the trigger time 
($T_0=$~2005 July 30.8322 UT). Although sparsely sampled, it appears that the 
$\rm B$ and $\rm V$ light-curves show similar behaviour up to 0.1~days post-burst. 
Near achromatic variability is clearly seen in the afterglow light-curves 
of all the pass bands. 
However, the light-curves do not display a correlated 
flaring behavior  observed in the early X-ray afterglow light-curve 
(Figures\,\ref{xrt_xmm} and \ref{compare}).
The values of early time flux decay indices using a single 
power-law fit ($f_\nu \propto t^{\alpha_1}$) to $\rm R$ and $\rm I$ data 
points between 0.01 to 0.1 day after the burst are $\alpha_1$ = -0.54$\pm$0.05 
and -0.66$\pm$0.11 respectively. Similar early time decay slopes were also 
obtained from the $\rm B$, $\rm V$, $\rm r'$ and $\rm i'$ 
afterglow light curves over a similar time scale 
but the $\rm R$ and $\rm I$ band light curves resulted in 
the best fits. The $\rm J, I$ and $\rm V$ light-curves, 
although sparsely sampled, show a bump followed by a considerable 
steepening at about 0.1~day (8.6~ks). The $\rm R$ and $\rm I$ light-curves 
from 0.1 day were fit by power-law index values ($f_\nu \propto 
t^{\alpha_2}$) $\alpha_2$ = -1.75$\pm$0.05 and -1.66$\pm$0.07 respectively.
Late time $V$ and $J$ afterglow light 
curves also show $\alpha_2 \sim$ -2 within the observing span.
Superimposed variability in the multi-band afterglow 
light-curves of GRB~050730 does not allow a reasonable fit for the generic 
broken power-law model (Beuermann et al. 1999). Therefore, we 
conclude that the early time light-curve decay slope $\alpha_1$ is -0.60$\pm$0.07 
and after a break time around 0.1 day, the weighted mean value of $\alpha_2$ is
-1.71$\pm$0.06 using $R$ and $I$ band data.

\begin{figure}[h]
\centering
\includegraphics[width=\columnwidth]{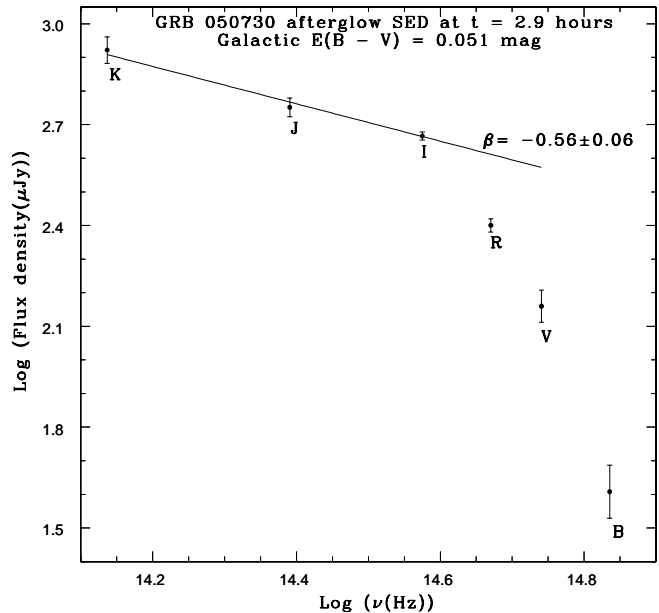}
\caption{$\rm NIR-$optical Spectral energy distribution ($\rm SED$) of 
GRB~050730 at $t$ = 2.9 hours (10.4 ks) after the burst. The magnitudes are
corrected for the galactic extinction of $E(B - V) = 0.05$ mag in the burst
direction. The epoch was chosen at the simultaneous $\rm VIJK$ observations using 
ANDICAM mounted at 1.3m telescope at CTIO. The effect of damped Ly-$\alpha$ due to 
high redshift (z $\sim$ 3.97) of the burst is clearly seen between $I$ and $R$ 
data points in form of considerable drop in the observed flux values.} 
\label{sed}
\end{figure}

\subsection{$\rm NIR$-Optical Spectral Energy Distribution}

A Spectral Energy Distribution (SED) of the afterglow of GRB~050730
was generated at 2.9~hours (10.4~ks) after the burst trigger.
This epoch was chosen to avail of simultaneous ANDICAM 
data in the $\rm V, I, J$ and $K$ pass-bands.
The SED from $\rm NIR$ to optical frequencies is presented in 
Figure\,\ref{sed}.
The reddening map provided by Schlegel, Finkbeiner \& Davis (1998) indicates 
a small value of $\rm E(B - V) = 0.05$ mag for the Galactic interstellar 
extinction towards the burst. We used the standard Galactic extinction 
reddening curve given by Mathis (1990) to convert apparent magnitudes into 
fluxes, with the effective wavelengths and normalizations from Bessel et al. 
(1998) for $\rm V, R, I, J, K$ pass-bands. Due to the high redshift 
(z $\sim$ 3.97) of the burst, the Ly-$\alpha$ break lies between $R$ and $I$ 
pass-bands, making $\rm NIR$ frequencies important to determine the correct spectral 
index of the burst. Since $R$ passband flux is also partially suppressed by the 
Ly-$\alpha$ break, we used only $\rm I, J, K$ data to determine the spectral 
index at the epoch of the $\rm SED$. If no spectral break occurs, the $\rm SED$ 
is generally described as a power law: $F(\nu) \propto \nu^{\beta}$, where 
$\beta$ is the spectral index. The derived spectral index in this way is $\beta$ = 
-0.56$\pm$0.06. Intermediate spectroscopic observations of the afterglow taken 
$\sim$ 3 hours after the burst by Starling et al. (2005) show negligible extinction 
in the host.

\begin{figure}[h]
\begin{center}
\includegraphics[width=\columnwidth]{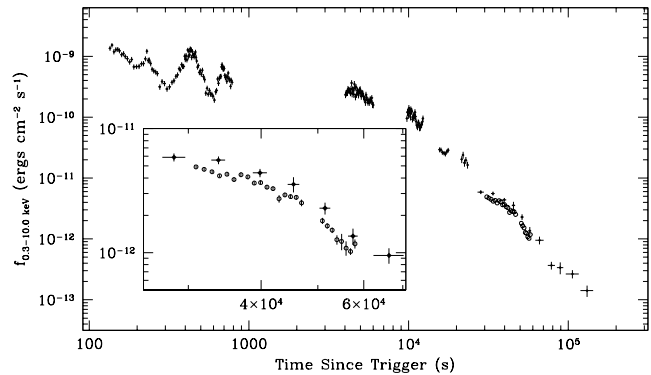}
\caption{The \textit{Swift}/XRT (closed circles) and \emph{XMM-Newton} (open circles) 
light-curve of GRB\,050730. Superimposed variability is clearly visible till 
the last epoch of the light-curve. Inset shows
the correlation between \it{Swift}/XRT and \emph{XMM-Newton} results.
\label{xrt_xmm}}
\end{center}
\end{figure}

\subsection {\textit{Swift} data analysis}  

\subsubsection{Light-curve Analysis}
 The {\it Swift}/XRT light-curve is complex with flares in the early WT data at 
$\sim$250~s, $\sim$420~s and $\sim$650~s (Figure\, \ref{xrt_xmm}, see also 
Starling et al. 2005). These early time flaring behaviors with rising or decaying
slope index values $\sim$ -6 are unusual and are not expected by simple synchrotron
fireball model (Sari, Piran \& Narayan 1998). A flare is also detected in PC data 
at $\approx$ 4500~s. The early data cannot be fit by a single power-law due 
to the flares. The light-curve after 15~ks can be fit to a power-law but
shows evidence of variability in some time intervals with an overall 
decay index -2.5$\pm$ 0.15, in agreement with Perri et al. (2005).
The \textit{Swift}/XRT light curve is presented together with the $\rm R$ and 
$\rm V$ band light curves in Figure~\ref{compare} for comparison purposes.

\begin{figure}[htbp]
\begin{center}
\psfrag{nH}[c]{\tiny $N_H$ $\times$ 10$^{22} $ cm$^{-2}$} 
\includegraphics[width=0.9\columnwidth]{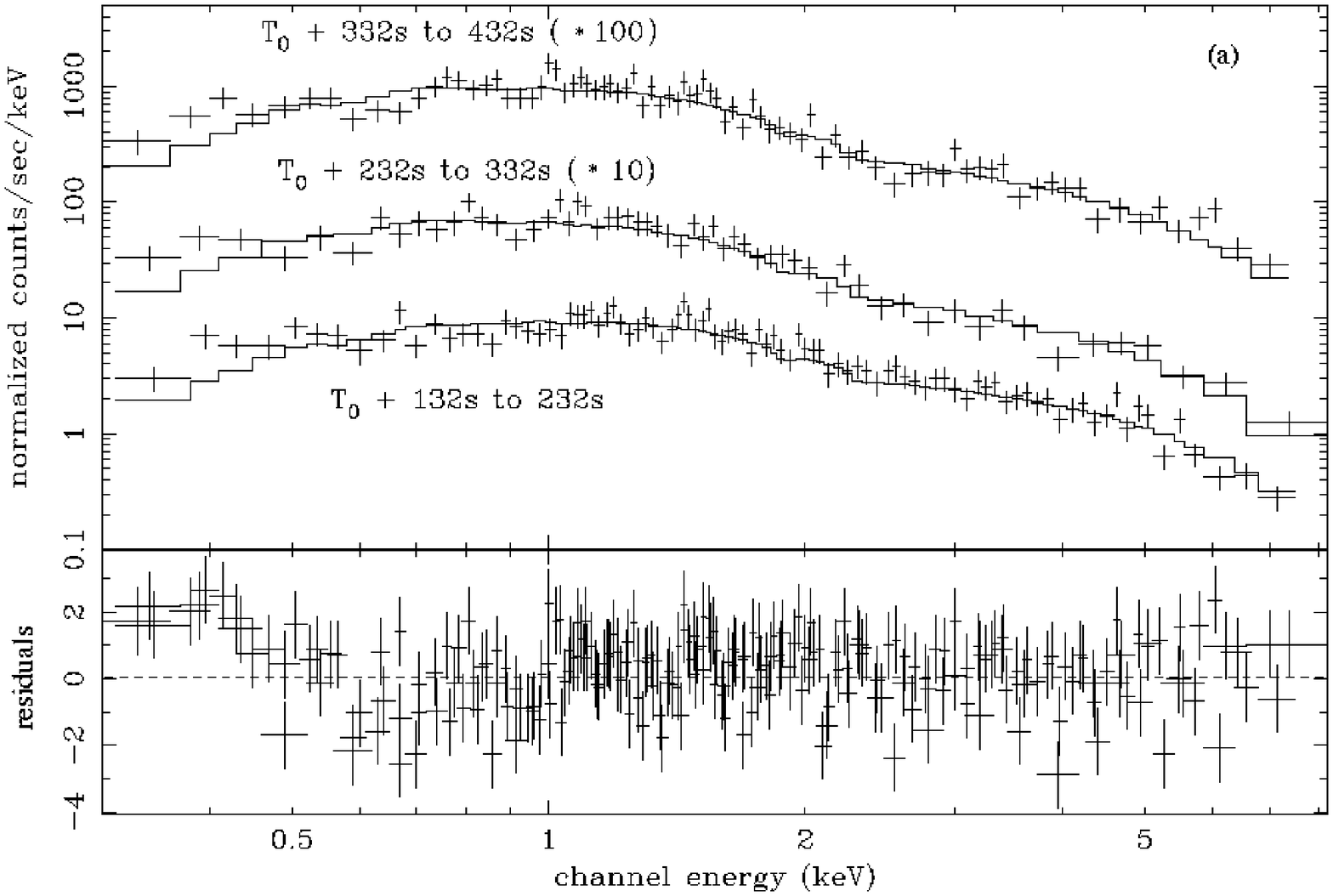}
\includegraphics[width=0.9\columnwidth]{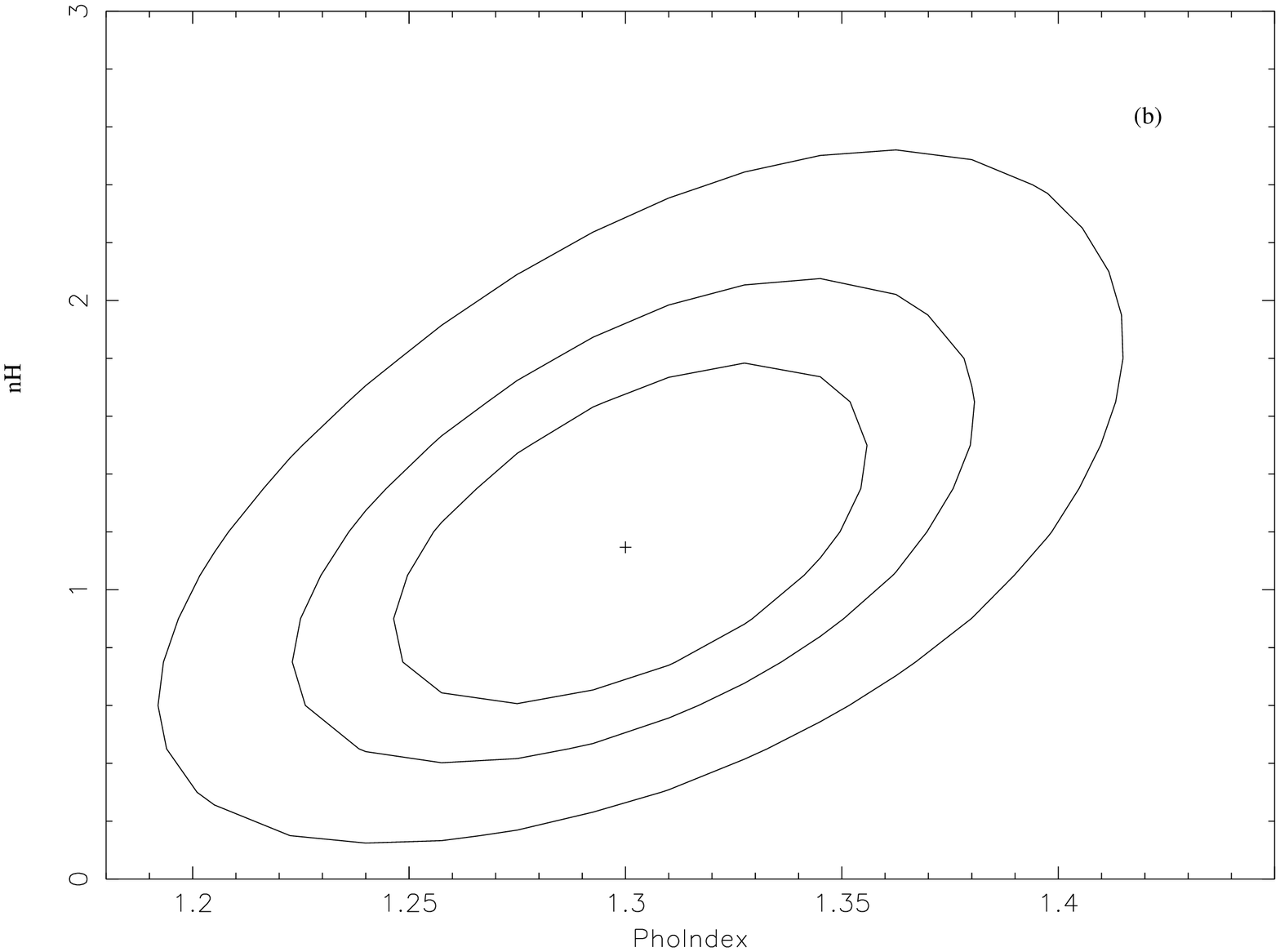}
\caption{(a) Simultaneous spectral fit for WT intervals from
T$_0+$132~s to T$_0+$432~s, best fit parameters are available in Table 4
(spectra are offset for presentation purposes)
and (b) Two dimensional confidence contours at 68.3\%, 90\% and 99\% for $N_H$ of 
the absorber component at the burst redshift and the photon index for the first interval.
\label{spec_WT}}
\end{center}
\end{figure}

\begin{figure}[htbp]
\begin{center}
\psfrag{nH}[c]{\tiny $N_H$ $\times$ 10$^{22} $ cm$^{-2}$}
\includegraphics[width=0.9\columnwidth]{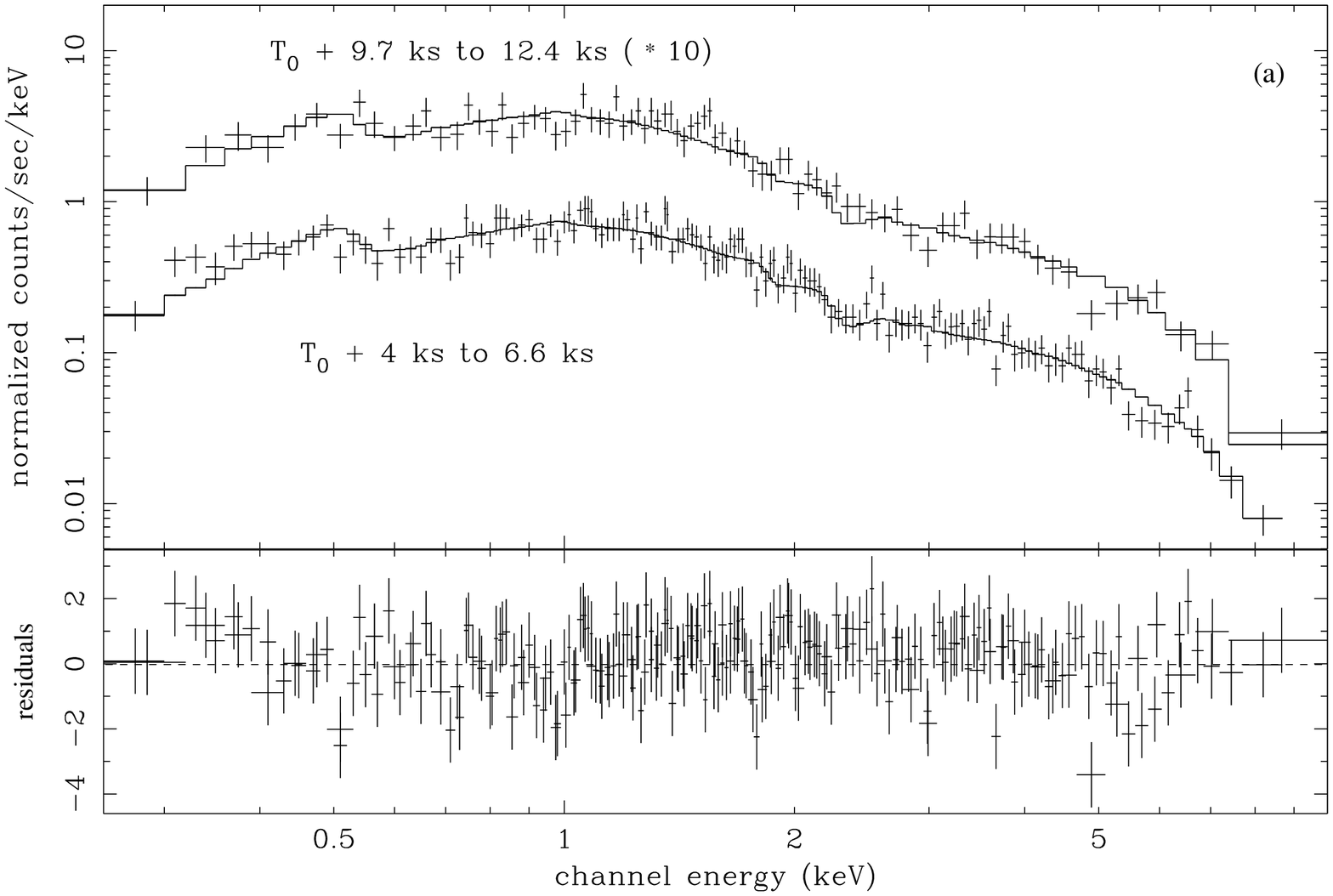}
\includegraphics[width=0.9\columnwidth]{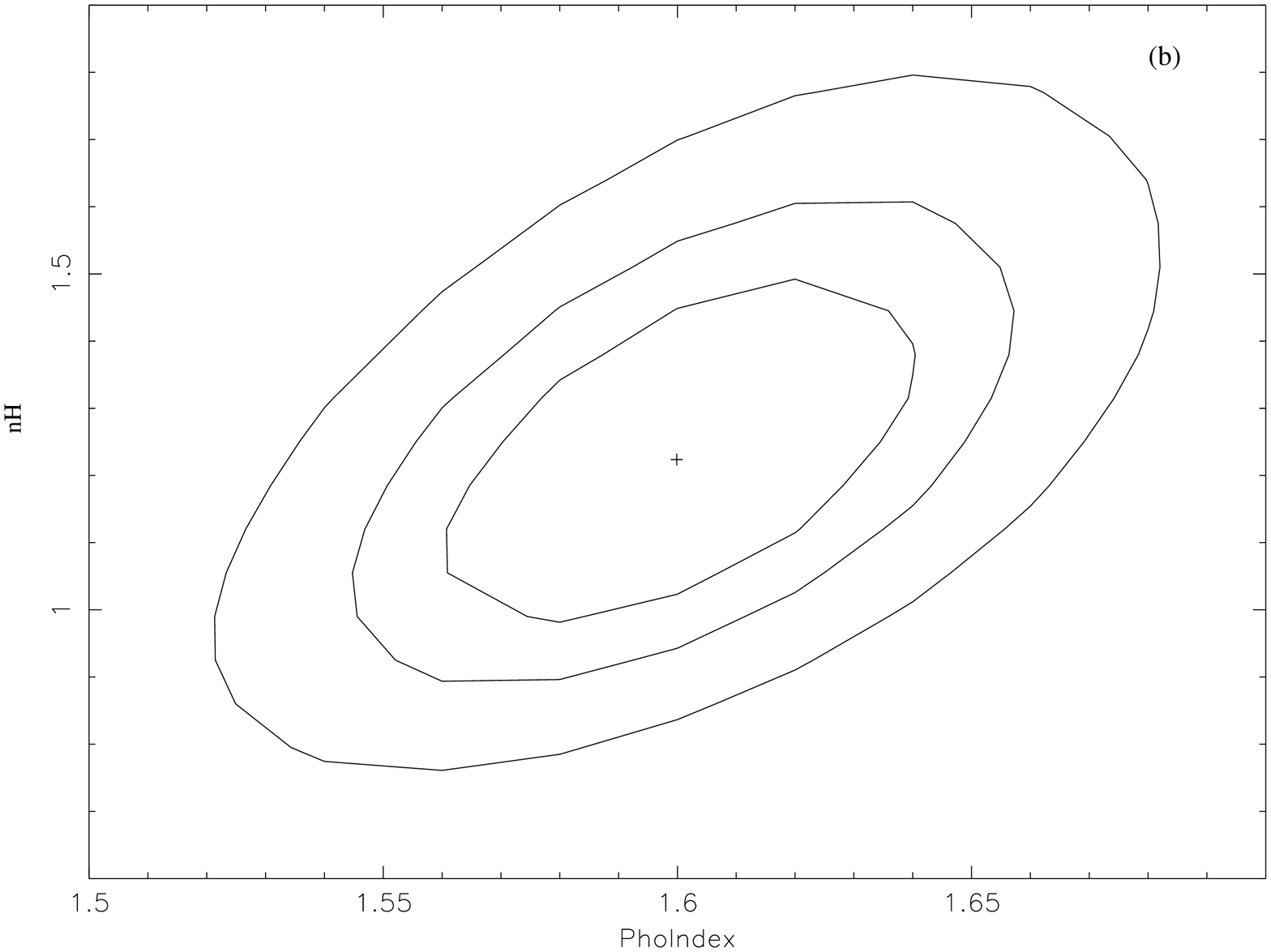}
\caption{(a) Simultaneous spectral fit for PC intervals T$_0+$4~ks to T$_0+$12.4~ks, 
best fit parameters are available in Table 4 
(spectra are offset for presentation purposes)
and (b) Two dimensional confidence contours 
at 68.3\%, 90\% and 99\% for $N_H$ of the absorber component at the 
burst redshift and 
the photon index for the first interval.
\label{spec_PC}}
\end{center}
\end{figure}

\subsubsection{Spectral Analysis}

Spectra were extracted from the \textit{Swift/}XRT data for 11 time 
intervals and the results are presented in Table 4. 
The spectra were fit with an absorbed power-law model including galactic 
column density, $N_H^{GAL}$, 3.2$\times10^{20}$ cm$^{-2}$ (Dickey \& Lockman 1990) and 
intrinsic absorption in the host at $z=$3.967. Several spectra were 
fit simultaneously to obtain an individual column density and to allow the power-law 
indices to vary independently (Table 4). A simultaneous spectral fit  
of the first three WT spectra from T$_0+$132~s to T$_0+$432~s and a contour plot 
of the column density at $z=$3.967 versus the photon index ($\Gamma$) of the first 
interval are shown in Figure\,\ref{spec_WT}. Combining these three intervals  
yields a column density of (11.4$\pm$0.7)$\times 10^{21}$\,cm$^{-2}$.
Only a 99\% confidence upper limit of  3.0$\times 10^{21}$\,cm$^{-2}$ 
could be obtained for the column density using the following three WT
spectra in the intervals from T$_0+$432~s to T$_0+$790~s (Table 4).

A simultaneous fit of the first two PC spectra from 
T$_0+$4~ks to T$_0+$12.4~ks and a contour plot of the column density at $z=$3.967 
versus the photon index of the first interval are shown in 
Figure\,\ref{spec_PC}. A tabulation of  all the spectral 
fits to the \textit{Swift}/XRT data and  intervals is available in Table 4.

The photon index evolves in afterglow from an early value of $\sim$1.30 
to $\sim$1.75 (or $\sim$1.9 using \emph {XMM-Newton} data (see section 3.4.2)). 
Starling et al. (2005) reported an excess column density in the first 
half of the WT observation and an abrupt change at T$_0+$500~s and an increase in the 
photon index consistent with these results.

It should be noted that the $\chi^2$/dof values in Table 4 for the WT data from 
T$_0$+132 to T$_0$+423~s indicate that the absorbed power-law model may not 
be the best model (Figure 6). A $\chi^2$/dof value of 310/228 is achieved by simultaneously 
fitting an absorbed power-law model to these data resulting in the large 
uncertainities in the intrinsic $N_H$ values. The data in the interval T$_0$+132 to 
T$_0$+423~s were also fit with a broken power-law model and an absorbed power-law with 
a black body component and the $\chi^2$/dof values were 264.2/227 and 265.2/226 respectively. 

In later epochs, T$_0$+432~s onwards, the $\chi^2$/dof values for the simultaneous fits are 
closer to unity. The absorbed power-law fit from T$_0$+432 to T$_0$+790~s yields $\chi^2$/dof 
of  198.9/184. The $\chi^2$/dof for the epoch T$_0$+4~ks to T$_0$+6.6~ks is 232/230 and the 
value for the epoch T$_0$+15.5~ks to T$_0$+140~ks the value is 99/102. The results for the 
absorbed power-law fits are presented in Table 4 for all epochs to allow a comparison 
between the early and late time \textit{Swift}/XRT spectra and the \emph{XMM-Newton} data.
The evolution of the power law index and the column density are presented with the 
X-ray, $\rm R$ and $\rm V$ light curves in Figure~\ref{compare}.

\begin{table*}[htdp]

{\bf Table 4.}~Spectral fit parameters from {\it Swift} and \emph{XMM-Newton} 
data analysis. The columns are the time interval over which the spectra 
were fit, the column density at the redshift of the host \textit{z}$=$3.967, the 
photon index, observed flux, unabsorbed flux, $\chi^2$ and degrees of freedom. The 
column density for the {\textit Swift/}XRT data was obtained by fitting the time 
intervals simultaneously while allowing the photon index to vary. Errors are quoted at 
the 90\% level for each parameter of interest.

\begin{center}
\begin{tabular}{|c|c|c|c|c|c|c|c|}
\hline
Time Interval & Column Density $N_{\rm H}$ & Power-law Index & f$_{obs}$ & f$_{unabs}$ & $\chi^2$ & d.o.f \\
(seconds since Trigger)& $\times10^{21}$ cm${-^2}$ &$\Gamma$ &
ergs cm$^{-2}$s$^{-1}$& ergs cm$^{-2}$s$^{-1}$ & & \\\hline \hline
$^\ddag$132$-$232 & & 1.30$\pm{0.07}$
& 9.7$\times$ 10$^{-10}$  & 10$\times$ 10$^{-10}$ & 120 & 92\\
232$-$332 & 11.4$^{+7.0}_{-7.0}$        & 1.7$\pm$0.13   &
4.4$\times$10$^{-10}$ & 5.1$\times$ 10$^{-10}$ & 80.6 & 58\\
332$-$432 &     & 1.50$\pm$0.09 &
7.3$\times$ 10$^{-10}$ & 7.9$\times$ 10$^{-10}$ & 107.6 & 79\\\hline
432$-$532 &  & 1.56$^{+0.10}_{-0.13}$ &
7.3$\times$ 10$^{-10}$ & 7.7$\times$ 10$^{-10}$ & 97.9 & 84 \\
532$-$632 &  3.0 $^\dagger$& 1.87$\pm$0.06 & 4.4$\times$ 10$^{-10}$&
4.8$\times$ 10$^{-10}$ & 96.5 & 94      \\
632$-$790 &     & 1.84$\pm$0.05 & 4.5$\times$ 10$^{-10}$ &
4.9$\times$ 10$^{-10}$ & 86.7 & 95      \\\hline
4$-$6.6~ks &    &       1.60$\pm{0.05}$ &
2.2$\times$ 10$^{-10}$ & 2.4$\times$ 10$^{-10}$ & 148.6 & 148\\
        &       12.2$^{+3.5}_{-3.0}$    &               &  &  & & \\
\vspace{-1.5em}  &      &       &       &       &       &\\
9.7$-$12.4~ks & & 1.73$\pm$0.07 &
1.0$\times$ 10$^{-10}$ & 1.1$\times$ 10$^{-10}$ & 89.9 & 111    \\\hline 
15.5$-$23.9~ks &        &       1.73$^{+0.10}_{-0.09}$
& 2.4$\times$ 10$^{-11}$& 2.8$\times$ 10$^{-11}$ & 34.6 & 39 \\
27$-$47~ks &    11.5$\pm6.0$&   1.84$^{+0.16}_{-0.13}$ &
4.5$\times$ 10$^{-12}$ & 4.8$\times$ 10$^{-12}$ & 34.1 & 41 \\
50$-$140~ks &  & 1.75$\pm0.14$ & 4.8$\times$ 10$^{-13}$ &
5.7$\times$ 10$^{-13}$ & 28 & 20 \\\hline \hline
$^\S$30$-$56~ks & 6.7$^{+1.9}_{-1.9}$ & 1.91$\pm0.03$ & 3.0$\times$ 10$^{-12}$ & 3.4$\times$ 10$^{-12}$ & 388.9 & 401 \\ 
30$-$40~ks & 7.1$^{+2.4}_{-2.4}$ & 1.87$\pm0.04$ & 4.1$\times$ 10$^{-12}$ & 4.6$\times$ 10$^{-12}$ & 256.9 & 302 \\ 
40$-$56~ks & 5.5$^{+3.0}_{-3.0}$ & 1.96$\pm0.04$ & 2.0$\times$ 10$^{-12}$ & 2.3$\times$ 10$^{-12}$ & 200.2 & 202 \\ \hline
\end{tabular}
\end{center}
\footnotesize{$^\ddag$ {\it Swift} epochs; $^\S$ \emph{XMM-Newton} epochs ;$^\dagger$ 99\% confidence 
Upper limit.}
\end{table*}

\subsection{\emph{XMM-Newton} data analysis}  

\subsubsection{Light-curve Analysis}

We have extracted background-subtracted light-curves in the 0.3-10.0 
keV energy band of the X-ray afterglow of GRB\,050730 using the 
circular source region of radius 20\arcsec\ shown in Figure~\ref{pn-image}.  
For the background region, an elliptical region 3 times the area of the 
source region and located at the same Y coordinate of the EPIC/pn CCD 
as the source region was used (also shown in Figure~\ref{pn-image}).  
The EPIC/pn and combined EPIC/MOS background-subtracted light-curves 
with time bins of 1 ks of the X-ray afterglow of GRB\,050730 are shown 
in Figure~\ref{xrt_xmm}.  The \emph{XMM-Newton} light-curve shows an
agreement with the {\it Swift} light-curve in the nature of decay as shown
in the inset of Figure~\ref{xrt_xmm}.

\subsubsection{Spectral Analysis}

\begin{figure}[htbp]
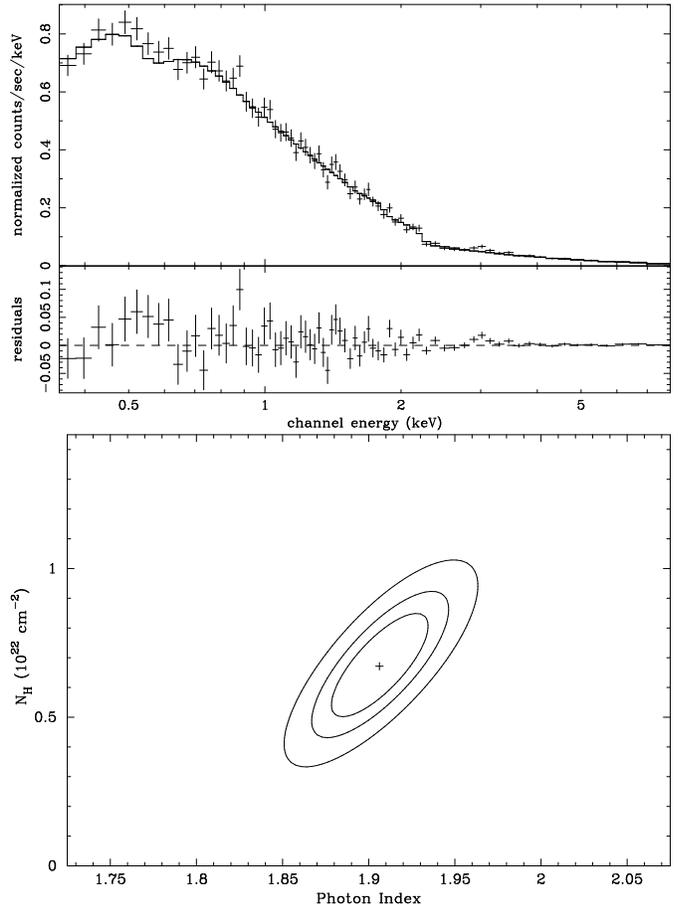

\begin{center}
\includegraphics[height=\columnwidth,angle=-90,clip=]{Spec_total.eps}
\includegraphics[height=\columnwidth,angle=-90,clip=]{Spec_total_cont.eps}
\caption{Top one shows the EPIC/pn spectrum of GRB\,050730 for the total exposure time, 
best fit parameters are available in Table 4 and the bottom plot showing 
Two-dimensional confidence contours at 68.3\%, 90\%, and 99\% for 
$N_{\rm H}$ of the absorber component at the burst redshift and 
$\Gamma$ of the spectral fit to the total exposure time EPIC/pn 
spectrum of GRB\,050730.}
\label{fig1}
\end{center}
\end{figure}

Source and background EPIC/MOS and EPIC/pn spectra of the X-ray afterglow 
of GRB\,050730 were extracted from the same spatial regions as those used 
for the light-curve extraction described above and shown in Figure~\ref{pn-image}.  
When the background spectrum is scaled down to the area of the source
spectrum, it contributes $\sim$1\% of the net
source count rate, i.e., the background contribution to the observed 
spectrum is almost negligible. The background-subtracted EPIC/pn spectrum of 
the X-ray afterglow of GRB\,050730 is shown in Figure~\ref{fig1}.  
The EPIC/MOS spectra (not shown here) show a similar spectral behavior, 
at lower signal-to-noise ratios than the EPIC/pn spectrum, so we
concentrate on the spectral analysis of this later one.  

\begin{figure}[htbp]
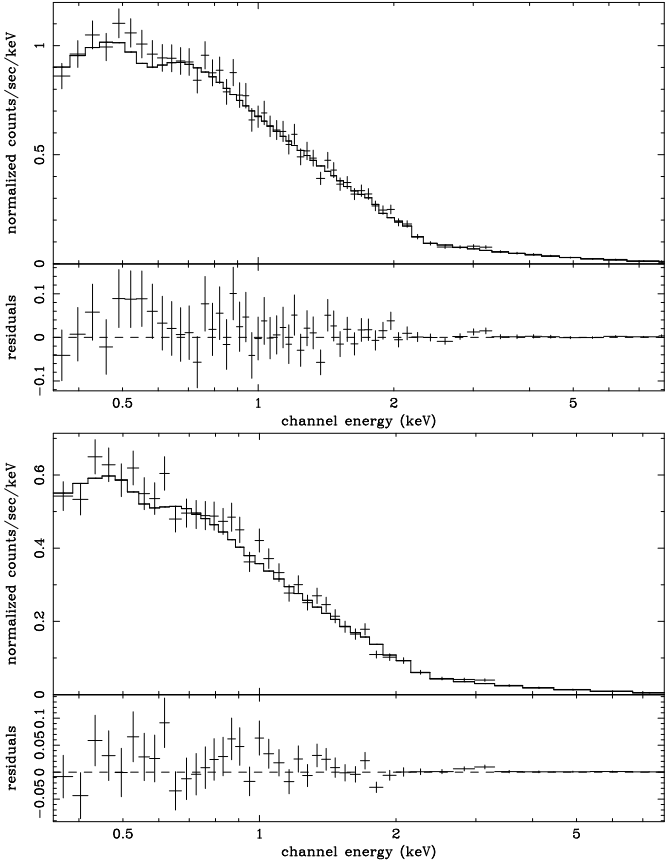

\begin{center}
\includegraphics[height=\columnwidth,angle=-90,clip=]{Spec_intA.eps}
\includegraphics[height=\columnwidth,angle=-90,clip=]{Spec_intB.eps}
\caption{The two figures are same as Figure~\ref{fig1} for the EPIC/pn spectrum for the 
first time segment (top) and for the last time segment (bottom).}
\label{fig2}
\end{center}
\end{figure}

The spectra were fit by an absorbed 
power-law model with two absorption components: an absorber component at the burst 
redshift, $z$=3.967, and a Galactic absorber component whose hydrogen column density, 
$N_{\rm H}^{\rm Gal}$, has been fixed at 3.2$\times$10$^{20}$cm$^{-2}$ 
(Dickey, \& Lockman, 1990). The same model was fit to the {\it{Swift}/XRT} 
spectra using XSPEC (Arnaud 1996).
The spectral fit was carried out by folding the absorbed 
power-law model spectrum through the EPIC/pn response matrix, and comparing the 
modeled spectrum to the observed EPIC/pn spectrum in the 0.35-8.0 keV energy 
range using the $\chi^2$ statistics. To ensure that the $\chi^2$ statistics can be used, 
the spectrum was binned to have a s/n $>$ 5 per spectral bin.  

The best-fit model, shown in Figure~\ref{fig1}, provides an excellent 
fit to the EPIC/pn spectrum, with a reduced $\chi^2$ close to unity, 
$\chi^2$/dof=0.98. The quality of the spectral fit is further illustrated by the 
plot of confidence contours as a function of $\Gamma$ and 
$N_{\rm H}$ shown in Figure~\ref{fig1}. The fitted model parameters  
with 90\% confidence level are listed in the last row of Table 4;  
the spectral photon index, $\Gamma$, is 1.91$\pm$0.03, and the hydrogen column 
density of the absorber component at the burst redshift is 
(6.7$\pm$1.9)$\times$10$^{21}$ cm$^{2}$. A look at the fit residuals in 
Figure~\ref{fig1} could suggest the presence of possible lines in the spectrum.  
We therefore add to the above model a thermal model component (the MEKAL model in 
XSPEC) and Gaussian lines at different energies, but in all cases the 
corresponding $\chi^2$/dof departed from unity. We confidently conclude that 
the X-ray emission of the afterglow of GRB\,050730 can be described by an 
absorbed power-law model with no significant contribution of thermal emission.  
The EPIC/pn spectrum of GRB\,050730 does not show any noticeable emission 
line and its spectral shape is suggestive of power-law models, representing 
synchrotron emission from a population of relativistic electrons. 

%

We have further searched for spectral variability throughout the \emph{XMM-Newton} 
EPIC/pn observation of GRB\,050730. The total useful exposure (after subtraction of 
periods of high-background) was divided into 2 segments including the starting 
8.2 ks and final 9.6 ks of the total exposure. The corresponding EPIC/pn spectra and 
best-fit models are shown in Figure~\ref{fig2}. The quality of these fits is not
better than the combined spectrum fit based on the value of $\chi^2$ tabulated
in Table 4. The other parameters of these best-fit models are also listed 
in Table 4. An inspection of these parameters suggests a marginal steepening 
of the spectral index.  

\section{Discussion}

\begin{figure}[htbp]
\begin{center}
\includegraphics[width=\columnwidth]{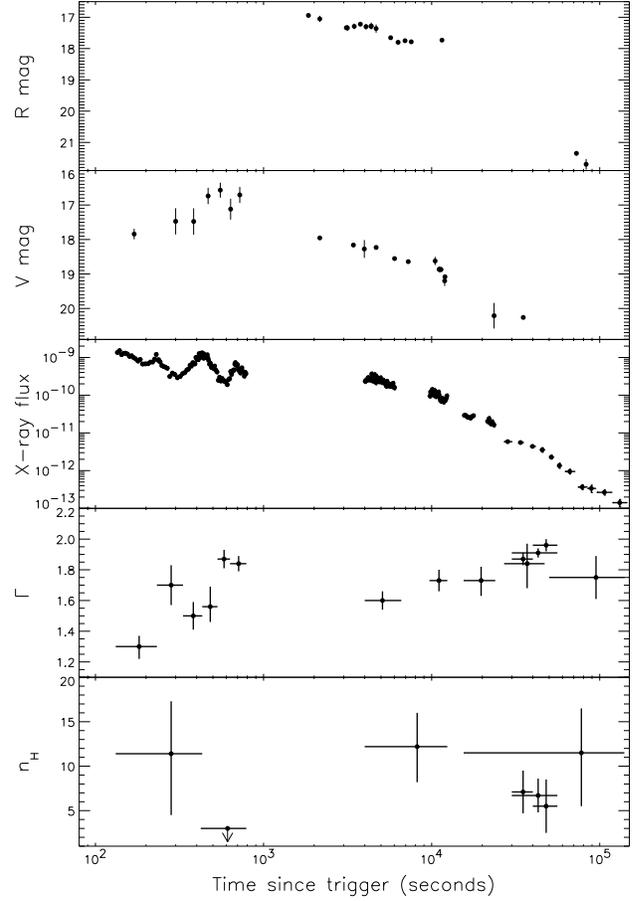}
\caption{The temporal evolution of parameters column density $N_H$ 
(in units of $\times$ 10$^{21} $ cm$^{-2}$) and photon index are shown, 
derived from the X-ray afterglow spectrum of GRB~050730. The X-ray afterglow 
light-curve along with the optical ones are also shown in the figure. The overall
nature of the light curves shows variability but the poor sampling of the optical
data does allow any strong correlation in the observed variabilities at the epoch
of observations from the earliest to the last common epochs. 
\label{compare}}
\end{center}
\end{figure}

We have presented early time optical, $\rm NIR$ photometry of the afterglow of 
GRB~ 050730 and millimeter observations. 
Optical afterglow  light-curves (Figure~\ref{optical} and \ref{compare}) show 
early time superimposed variability similar to that observed 
in the case of GRB 000301C (Sagar et al. 2000) and GRB 021004 (Pandey et al. 2003). 
X-ray afterglow observations of GRB~050730 from {\it Swift} and \emph{XMM-Newton} 
(Figure~\ref{xrt_xmm}) also show a variability until 
the last epoch of observations (Figures 5 and 10). The 
observed X-ray and optical afterglow variability seems to be uncorrelated in 
very early phases (t $<$ 0.01 day) of the light-curves. 
The nature of the early X-ray flaring and its correlation 
with BAT light-curves has been discussed by (Burrows et al. 2006; 
Nousek et al. 2006) and may indicate an ongoing central engine activity superimposed 
on a slowly decaying initial afterglow phase. The observed flaring behavior can be 
explained in terms of one of the theoretical models proposed for such 
variabilities (King et al. 2005, Perna et al. 2005, Proga \& Zhang 2006) for the 
massive star origin of the burst.

X-ray afterglow data from \emph{XMM-Newton} taken $\sim$ 25 ks after the burst 
show a decay similar to that of the {\it Swift/}XRT light-curve as shown in the
inset of Figure~\ref{xrt_xmm}. 
The {\it Swift/}XRT temporal and spectral analysis are in 
agreement with those of the \emph{XMM-Newton} data (Table 4). The X-ray 
afterglow light-curve 15~ks after the burst shows an overall temporal flux decay 
index of -2.5$\pm$0.15, steeper than the late time optical temporal decay index
and comparison of the X-ray and optical light-curves can be seen in 
Figure~\ref{compare}.
Temporal evolution of $\Gamma$ is clearly seen from the first to the last epoch of 
observations and the value of column density is overall similar except for a sudden 
drop around 500 seconds after the burst as initially reported by Starling et al. (2005).
This might resemble the variable column density seen in the case of 
prompt emission of GRB 000528 (Frontera et al. 2004).

The averaged temporal flux decay indices $\alpha_1 = -0.60\pm0.07$ and 
$\alpha_2 = -1.71\pm0.06$ are derived based on a considerable steepening in 
the $R$ and $I$ data points around 0.1 day (8.6 ks) after the burst. The {\textit Swift} 
light-curve after 0.1 day can be fit by a power-law decay 
index $\alpha_2 = -2.5\pm 0.15$, steeper than that derived from optical frequencies. 
The spectral index $\beta = -0.56\pm0.06$ is
determined from $I, J$ and $K$ pass-band observations at $\sim$ 10.4 ks after 
the burst. The value of $\beta$ derived from the X-ray photon index value (Table 4) 
at a similar epoch is $-0.73\pm0.07$, statistically in agreement with that derived from 
optical-$\rm NIR$ frequencies, and indicates that the cooling break frequency 
$\nu_c$ does not lie between optical and X-ray frequencies at the given epoch. 
The derived values of $\alpha_2$ and $\beta$ from optical and X-ray frequencies 
in the light of slow-cooling $\rm ISM$ jet model predictions (Sari, Piran \& 
Halpern 1999, Rhoads 1999), rules out the  possibility of $\nu_c$ below $\rm NIR$ 
frequencies. The location of $\nu_c$ above observed X-ray frequencies at the 
time of the $\rm SED$ gives the electron energy index $p \sim 2.3$, in agreement 
with $\alpha_2$ derived from the X-ray light-curve. Such a high value of 
$\nu_c$ at the epoch of the observations indicates a relatively low values of 
the post shock magnetic field energy $\epsilon_B$ and ambient density (Sari,
Piran \& Halpern 1999). The location of $\nu_c$ above observed X-ray frequencies 
around 0.1 day (8.6 ks) after the burst, and no considerable evolution of 
X-ray spectral index (see Table 4) from 0.1 until 1.5 day (8.6 ks to 130 ks) 
after the burst, indicates that the observed break in the optical light-curves 
is a jet-break. Such an early jet-break is also reported in the case of another 
high redshift burst, GRB 050319 (Cusumano et al. 2006) but not in the case of the 
highest redshift burst, GRB 050904 (Tagliaferri et al. 2005) known as of today.  

The flatter value of $\alpha_2$ at optical frequencies with 
respect to the X-ray regime is inconsistent with the standard fireball model predictions 
(Sari, Piran \& Halpern 1999). The hypothesis of a possible underlying host galaxy 
or contribution from an associated supernova to the observed shallower value of 
$\alpha_2$ at optical frequencies can be ruled out considering the high 
redshift of the burst. Apart from an observed variability around 0.1 day 
after the burst, the $R$ and $I$ afterglow late time light-curves do not show 
further variability. Other plausible 
explanations for the observed shallower $\alpha_2$ at optical frequencies are in 
terms of modified afterglow models: refreshed shocks or fluctuations in the 
external media. In refreshed shock models the fluctuations in the observed flux 
are expected at both the frequencies, but lack of X-ray observations at later 
epochs does not allow us to constrain the observed flatness at optical frequencies 
in terms of the model. In the case of the model where the fireball encounters
regions of enhanced density (Lazzati et al. 2002; Nakar, Granot 
\& piran 2003), afterglow flux is supposed to be considerably dependent on the 
external density for the frequencies below $\nu_c$. In the present case, the location 
of $\nu_c$ above observed X-ray frequencies at very early epochs ($\sim$ 0.1 day) 
does not allow this interpretation of the observed flatness at optical frequencies.
The possibility of the two component jet model (Berger et al. 2003) can also not 
be ruled out for the observed discrepancy between the $\alpha_2$ values at the
two frequencies although at radio frequencies no significant observations were found
(van der Horst \& Rol 2005a,b). The sparse temporal coverage and the 
lack of observations 3 days after the burst however, make the various explanations for 
the observed discrepancy between the $\alpha_2$ values at optical and X-ray frequencies
indistinguishable.

The interpretation of the observed steepening around 0.1 days as a jet-break in terms 
of the $\rm ISM$ jet model show $\alpha_2 \sim p$, but the observed value of 
$\alpha_1$ is flatter in comparison with the closure relation $\alpha_1 = 3\beta/2$. 
The observed early-time superimposed variability in the form of a flatter value of 
$\alpha_1$ might imply a set of energy-injection episodes 
(Zhang \& M\'{e}sz\'{a}ros 2002) followed by late-time central engine activity.
For a Poynting-flux-dominated continuous energy injection, 
Zhang \& M\'{e}sz\'{a}ros (2002)
assume that the source luminosity $L(t)\propto t^{q}$, where $q$ is the 
temporal index, and influences the observed light-curve through energy injections
for $q > -1$. The value of $q$ is related to the observed $\alpha$ and $\beta$ 
with a closure relation as long as the observed frequencies are in same spectral
region. In the case of $\nu_c$ being above X-ray frequencies, $\alpha, \beta$ and $q$
are related as $\alpha$ = $(1 - q/2)\beta + q + 1$ (Zhang \& M\'{e}sz\'{a}ros 2002).
The fact that the observed values of $\alpha_1$ and $\beta$ are in agreement with the above
closure relation with $q > -1$, supports the explanation of the observed flatter
value of $\alpha_1$ being due to the early-time light-curve being dominated by energy 
injection episodes. However, in order to understand the observed variability in terms of the
energy injection episodes, detailed modeling is required as in the 
case of GRB 010222 (Bj\"ornsson et al. 2002) and GRB 021004 
(de Ugarte Postigo et al. 2005). 

The millimeter detection around 3 days after the burst and an 
upper limit around 5 days after the burst (see Table 1) puts an important 
constraint on the temporal decay of the afterglow. Comparison of the millimeter and 
optical flux at the similar epochs in terms of a simple $\rm ISM$ jet model shows an 
order of $\sim$ 2 magnitude excess emission at millimeter frequencies with respect
to predictions. The expected millimeter flux was calculated using the value of the maximum 
synchrotron frequency $\nu_m$ in millimeter frequencies at the epoch of observations, 
the self absorption break frequency $\nu_a$ in radio frequencies, $\nu_c$ above X-ray 
frequencies, and the derived values of the jet-break time and $p$ were as discussed above.
The derived excess may imply the presence of variability in the millimeter regime
and possibly that energy injection contributes significantly at millimeter 
frequencies. The possibility of a millimeter bright host galaxy at this high 
redshift is ruled out on the basis of upper limits obtained from our 
further monitoring of the field around 160 days after the burst. 

The observed fluence of 4.4$\times 10^{-6}$erg/cm$^2$ in the energy band 15 -- 350 KeV 
with the measured redshift z = 3.967$\pm$0.001 implies an isotropic 
equivalent energy release $E_{\rm iso,\gamma}\sim 2.3 \times 10^{53}$~erg for 
$H_0$ = 65 km/s/Mpc in a $\Omega_0$ = 0.3 and $\Lambda_0$ = 0.7 cosmological 
model with cosmological K-correction (Bloom et al.\ 2001). 
If we take the observed break time of $0.1$~days as the jet-break 
time and an assumed value of $\gamma-$ray efficiency $\eta_{\gamma}=0.2$, this 
leads to an estimated jet half-opening angle $\theta \sim 1.3 \times n^{1/8}$ 
degree, where $n$ 
is the particle density of the ambient media.
The total $\gamma-$ray energy output in the jet then works out to be 
$E_\gamma \sim 5.7\times 10^{49}$~erg which lies at the low end of the 
observed beaming corrected $E_\gamma$ for long duration GRBs (Frail et al. 2001) 
and appears to be underluminous. 
A possible explanation for the below average $E_\gamma$ 
may be either considerable energy at lower frequencies in form 
of energy injection episodes or afterglow kinetic energy. 
The observed early jet-break time implies that this burst was viewed almost pole-on.
Although the multi-wavelength observations of GRB~050730 
have uncovered the peculiar nature of the afterglow, the gaps in the light-curves 
make it impossible to distinguish between several possible scenarios.

\section{Conclusion}

Multi-wavelength observations of the afterglow of GRB~050730 from millimeter
to $\rm NIR-$optical frequencies are used to analyze the burst properties. The unusual
nature in form of superimposed variability is seen in the light-curves 
from millimeter to X-rays. The derived values of the photon indices from the
{\it Swift} and \emph{XMM-Newton} data analysis along with the optical-$\rm NIR$ $\beta$ 
constrain the value of $p \sim 2.3$ and the location of the cooling break to be above 
the observed frequencies. Model predicted flux at millimeter frequencies shows an excess
in the observed millimeter fluxes around 3 days after the burst, indicating
possible variability at these frequencies too. The value of $\alpha_1$ derived from 
optical observations suggests the early time energy injections. The derived 
value of the jet-break time shows the burst to be under-luminous on the basis of the derived 
beaming corrected $E_\gamma$. Detailed modeling is encouraged to understand the nature 
of the observed variability and the missing energy in the form of possible energy 
injection episodes in this case.

The importance of early time multi-wavelength $\rm NIR-$optical observations is obvious 
in the case of the high redshift GRB~050730. The observed early time ($\sim$ 0.1 day in 
the observer's frame) break in the optical light-curves of this high redshift event, 
interpreted in terms of a jet-break, is not common in comparison with other well observed 
afterglows at lower redshifts. In future, X-ray afterglow observations at late 
phases will be essential in order to understand these primordial energetic explosions.

\section*{Acknowledgments}
This research has made use of data obtained through the High Energy 
Astrophysics Science Archive Research Center On line Service, provided by the 
NASA/Goddard Space Flight Center. S.B.P. acknowledges MAE-AECI grant for 
this work. J.G. and J.A.C. are researchers of the programme {\em Ram\'on y Cajal} 
funded jointly by the Spanish Ministerio de Educaci\'on y Ciencia (former Ministerio de
Ciencia y Tecnolog\'{\i}a) and Universidad de Ja\'en. The authors also
acknowledge support by DGI of the Spanish Ministerio de Educaci\'on y Ciencia 
under grants AYA2004-07171-C02-02, FEDER funds and Plan Andaluz de Investigaci\'on 
of Junta de Andaluc\'{\i}a as research group FQM322. M.A.G.\ acknowledges support 
by the Spanish program Ram\'on y Cajal. SMB acknowledges the support of the European Union 
through a Marie Curie Intra-European Fellowship within the Sixth Framework Program.
Millimeter observations from IRAM Plateau de Bure Interferometer are acknowledged. IRAM is 
supported by INSU/CNRS (France), MPG (Germany) and ING (Spain). The Liverpool Telescope is 
operated on the island of La Palma by Liverpool John Moores University at the 
Observatorio del Roque de los Muchachos of the Instituto de Astrof\'isica de Canarias.
Publicly available {\it Swift}/XRT and UVOT data and Leicester University 
HelpDesk are also acknowledged. We are also thankful to the anonymous referee for the constructive comments. 


\end{document}